\documentclass[a4paper,twocolumn,english,showpacs,nofootinbib,nobibnotes,aps,pra,floatfix]{revtex4-2}
\usepackage{graphicx}
\usepackage{bm}
\usepackage{amssymb} 
\usepackage{amsmath} 
\usepackage{amsthm}
\usepackage{hyperref}
\usepackage[utf8]{inputenc}
\hypersetup{
     colorlinks=true,      
    linkcolor=blue,        
    citecolor=blue,
    filecolor=magenta, 
    urlcolor=black          
}
\usepackage{array, boldline, makecell, booktabs, amsmath}

\usepackage{mathbbol}

\begin{document}

\title{Characterizing crosstalk of superconducting transmon processors}

\author{Andreas Ketterer}
\email{andreas.ketterer@iaf.fraunhofer.de}
\affiliation{Fraunhofer Institute for Applied Solid State Physics IAF, Tullastr.~72, 79108~Freiburg, Germany}
\author{Thomas Wellens}
\email{thomas.wellens@iaf.fraunhofer.de}
\affiliation{Fraunhofer Institute for Applied Solid State Physics IAF, Tullastr.~72, 79108~Freiburg, Germany}

\begin{abstract}
 Currently available quantum computing hardware based on superconducting transmon architectures realizes networks of hundreds of qubits with the possibility of controlled nearest-neighbor interactions.  However, the inherent noise and decoherence effects of such quantum chips considerably alter basic gate operations and lead to imperfect outputs of the targeted quantum computations. In this work, we focus on the characterization of crosstalk effects which manifest themselves in correlations between simultaneously executed quantum gates on neighboring qubits. After a short explanation of the physical origin of such correlations, we show how to efficiently and systematically characterize the magnitude of such crosstalk effects on an entire quantum chip using the randomized benchmarking protocol. We demonstrate the introduced protocol by running it on real quantum hardware provided by IBM observing significant alterations in gate fidelities due to crosstalk. Lastly, we use the gained information in order to propose more accurate means to simulate noisy quantum hardware by devising an appropriate crosstalk-aware noise model.
\end{abstract}

\maketitle

\section{Introduction} 
Superconducting transmon qubits are among the most promising qubit architectures as they allow to engineer quantum chips supporting networks of qubits of ever-increasing size with controlled nearest neighbor interactions~\cite{Arute:2019aa,Kjaergaard:2020aa,RevModPhys.93.025005,Jurcevic_2021,Wendin_2017}. Such networks of qubits are of particular interest as they allow for the implementation of quantum error correcting codes, such as the surface code or variants of it, which enable the realization of fault-tolerant quantum computations once the individual single- and two-qubit gate fidelities surpass the limits provided by the threshold theorem~\cite{10.1145/258533.258579,NielsenChuang,PhysRevA.80.052312,PhysRevX.10.011022}. A precondition for the threshold theorem to hold is, however, that errors which occur on the quantum chip are of a particular type, e.g., bit- and phase-flip errors of individual qubits or combinations thereof. If this cannot be ensured, e.g., if errors are of more complicated global type, the threshold theorem might not hold and, consequently, a successful implementation of quantum error correction is jeopardised.

Problematic processes that distort the performance of a certain number of qubits and occur frequently on quantum chips based on superconducting transmon architectures are, for instance, the occurance of two-level system (TLS) defects~\cite{Lisenfeld:2019vj,PhysRevB.92.035442}, energetic impacts from cosmic rays that ionize the substrate and lead to radiating high energy phonons~\cite{McEwen:2022ww,2022arXiv221004780T}, or crosstalk caused by frequency collisions among neighboring qubits~\cite{PhysRevA.102.042605,Hertzberg:2021te,PhysRevResearch.4.023079}. In general, crosstalk refers to errors that occur during the execution of gate operations on specific subsets of qubits and influence the performance of other qubits located in the vicinity of the latter. Frequency collisions, i.e., degeneracies of transition frequencies of neighboring transmon qubits, are one possible origin of crosstalk effects as they can lead to undesired resonances during the application of microwave pulses which realize the individual gate operations. 

Frequency collisions can occur in various forms and on different subsets of qubits depending on their physical origin. For instance, they can lead to undesired
excitations of neighboring qubits which 
may
lead to so-called leakage, i.e., population of higher energy levels outside of the computational subspace. The latter are particularly problematic because populations of higher energy levels permanently damage the respective qubits and thus influence the result of the executed computation. 
In other cases, frequency collisions can 
activate
couplings of neighoring qubits which have an undesired entangling effect. In general, the interactions induced by frequency collisions can be grouped in different categories depending on  whether they involve single- or  two-qubit gates and to which qubits they are applied to~\cite{Hertzberg:2021te}. 

In the present work, we investigate in detail crosstalk effects caused by frequency collisions on superconducting transmon processors. In this respect, we first discuss the physical origin of these effects and show that they manifest themselves in correlations between simultaneously executed quantum gates on neighboring qubits. Further on, we present a protocol which allows for a systematic and efficient characterization of the magnitude of such crosstalk effects on an entire quantum chip using the randomized benchmarking protocol. We demonstrate the introduced protocol by running it on real quantum hardware provided by IBM, observing significant alterations in gate fidelities. Lastly, we use the gained information in order to propose more accurate means to simulate noisy quantum hardware by devising an appropriate crosstalk-aware noise model.

\begin{figure*}[t!]
    \centering
    \includegraphics[width=1.95\columnwidth]{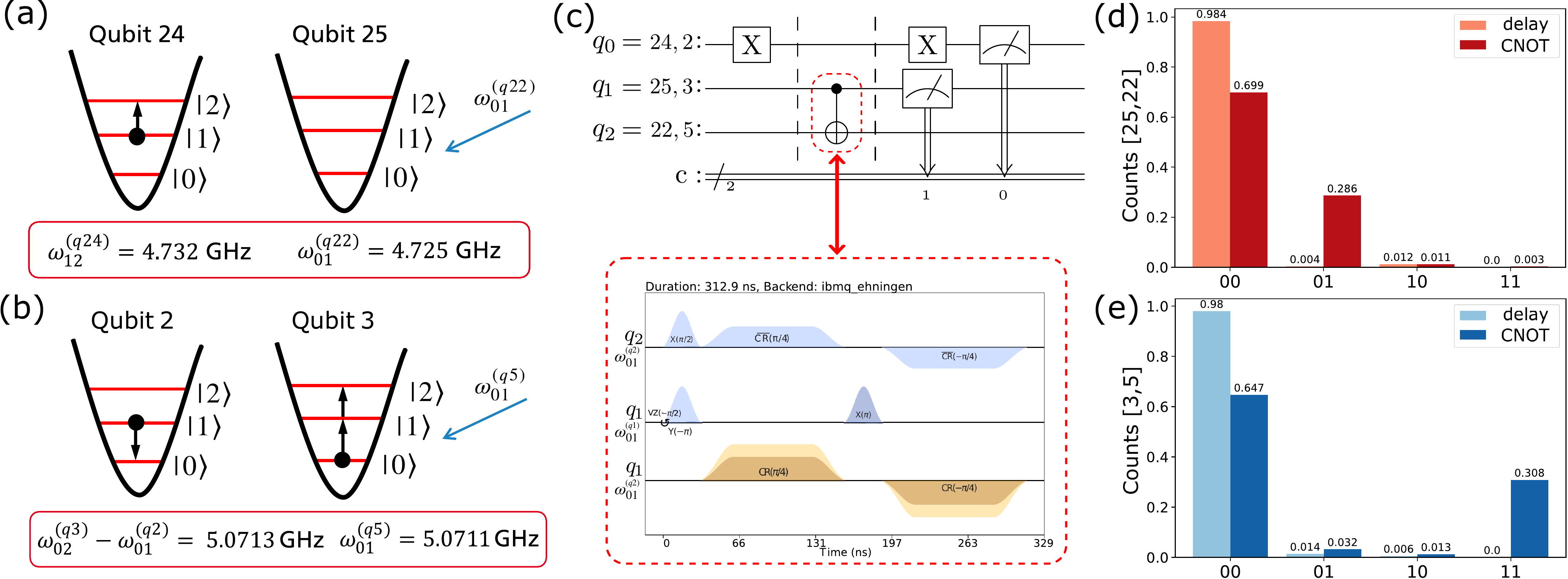}
    \caption{(a,b) Representation of two frequency collisions on 
   \textit{ibmq\_ehningen}
   that lead to resonances during the application of cross-resonance (CR) pulses. The transmon qubits are shown pictorially with their three lowest energy levels, and the blue arrows indicate the application of a CR-pulse to qubit 25, respectively 3. (c) Circuit used to drive the resonance shown in (a) and (b). The red dashed box depicts the pulse schedule of the CNOT gate involving the CR pulses shown in yellow. Each line corresponds to a channel where a pulse with frequency $\omega^{(q)}_{01}$ is applied to qubit $q$ and shaded colors indicate phases of the respective microwave pulses. (d,e) Histograms of the outcomes obtained from this circuit run on the two indicated sets of qubits (dark red, respectively blue). For comparison, we also measured the same circuits where, instead of the CNOT gate, a simple delay of the same duration is implemented (light red, respectively blue). 
   Since both outcomes should be identical in the absence of errors, the observed difference clearly demonstrates the influence of crosstalk in both cases (d) and (e).  
    }
    \label{fig_1}
\end{figure*}

The paper is organized as follows. In Sec.~\ref{sec:CrosstalkEffects}, we discuss the physical origin of the crosstalk effects under consideration and discuss their impact. We then continue in Sec.~\ref{sec:CrosstalkChar} by presenting the protocol for the systematic crosstalk characterization and present results obtained from real quantum hardware. In Sec.~\ref{sec:CrosstalkNoiseModel}, we present a crosstalk-aware noise model and demonstrate its performance in practice. Lastly, we conclude in Sec.~\ref{sec:Conclusions} and give a short outlook.

\section{System configurations and identification of crosstalk effects}\label{sec:CrosstalkEffects}

Transmons are engineered superconducting circuits realizing quantized anharmonic charge oscillators. Their lowest two energy levels are used as qubit states~\cite{PhysRevA.76.042319,PhysRevB.77.180502}. Among the various existing transmon architectures, we will focus on so-called fixed-frequency lattices, where each qubit exhibits a fixed transition frequency referring to the lowest energy transition and a corresponding anharmonicity that characterizes its nonlinearity. The latter is often  quantified through the difference between the $|0\rangle \rightarrow |1\rangle$ and $|1\rangle \rightarrow |2\rangle$ transition frequency, i.e., the lowest and next higher energy transition. Further on, transmon qubits can be manipulated by 
passing in engineered microwave pulses through transmission lines,
thereby enabling either single-qubit rotations or couplings between neighboring qubits allowing for the realization of entangling gates~\cite{PhysRevA.93.012301,PhysRevB.81.134507}.  Note that alternative transmon architectures allow for tunable transition frequencies which can be exploited to realize two-qubit gates based on tunable couplings among qubits~\cite{PhysRevApplied.8.044003,PhysRevApplied.10.054062}. However, in this work we will focus only on fixed frequency architectures. 

A possible origin of crosstalk effects on quantum chips based on superconducting transmon qubits is the collision of transition frequencies among neighboring qubits. The latter may lead to undesired driving of individual qubits or induce interactions between qubits and thus can severely falsify the implemented quantum computation. One way to identify such frequency collisions is to analyse the values of neighboring qubits' transition frequencies, as well as their  anharmonicities, and to search for resonances that can potentially be driven by microwave pulses of specific single- and two-qubit gate operations. For instance, the simplest example of such a resonance is given when the qubit-transition frequencies of two neighboring qubits, A and B, approximately coincide, i.e., $\omega^{(A)}_{01}\approx\omega^{(B)}_{01}$. In this case, single-qubit gates applied to qubit A can also drive the transition of qubit B and vice versa.  More complex resonances might involve also transitions to higher energy levels, e.g. $|2\rangle$, and thus lead to leakage from the qubit subspace $|0/1\rangle$. The latter are particularly problematic as the population of higher excited states cannot be undone by unitary transformations in the qubit subspace and thus permanently damage the performance of the involved qubits. A selection of possible frequency collisions is summarized in Table~1 of Ref.~\cite{Hertzberg:2021te}.

A frequency collision of the just discussed type is, however, rather unlikely because the degeneracy of qubit transition frequencies of direct neighbors can be largely avoided by appropriate chip engineering. In contrast, frequency collisions between next-nearest neighbor qubits are more difficult to avoid. The latter pose a particular problem during the application of two-qubit gates which might 
suffer from
such collisions and hit unwanted resonances. On superconducting hardware by IBM, e.g., on the Falcon r5.11 processor investigated later on, the only two-qubit basis gate is the CNOT gate, which is realized either by implementing a cross-resonance gate in combination with additional single qubit rotations \cite{PhysRevB.81.134507,PhysRevA.93.060302}, or by a so-called direct $C_X$ gate~\cite{PhysRevLett.127.130501,Jurcevic_2021}. In both cases, an interaction between the involved qubits, A and B, is induced by applying to one designated qubit, say qubit A, a microwave pulse whose frequency $\omega^{(B)}_{01}$ is chosen according to the respective other qubit's energy transition. Overall, this leads to a native CNOT gate that is controlled by qubit $A$ and whose control and target can be exchanged by applying additional Hadamard transformations to the control 
and target
qubit. 

We isolated two examples of frequency collisions that lead to unwanted resonances during the application of CNOT gates in case of the IBMQ system \textit{ibmq\_ehningen}, which is a $27$ qubit Falcon r5.11 transmon processor with heavy-hexagon topology and fixed frequency architecture (see Fig.~\ref{fig_2}). To demonstrate the severeness of these two collisions, we ran tailored experiments in order drive the respective resonances on purpose. In both cases, the experiments involve three qubits, one of which is prepared in state $|1\rangle$ followed by the application of a CNOT gate on 
a neighbouring pair og qubits
(see Fig.~\ref{fig_1}(c)). In doing so, the cross-resonance pulse underlying the CNOT gate drives one of the two resonances shown in Fig.~\ref{fig_1}, depending on which qubits are chosen.

In the first case, we focus on qubit 24 which, due to the application of the CNOT gate to the qubits 25 and 22, is partially excited from the state $|1\rangle$ to the next higher energy state $|2\rangle$. Consequently, another application of a Pauli $X$ gate will not be able to map all the qubit's population back to the ground state. The remaining population in state $|2\rangle$ will then be displayed as residual population in state $|1\rangle$, as the standard qubit readout implemented on the hardware is insensitive to measurement events resulting from the state $|2\rangle$. We measured this residual population of qubit 24 and present the result in Fig.~\ref{fig_1}(d). For comparison, we additionally ran the same circuit where the CNOT gate is replaced by a delay of the same duration as the CNOT gate time. 

The second collision, shown in Fig.\ref{fig_1}(b), is more subtle, as it leads to a resonance that induces a coupling between qubit 2 and 3, where one qubit is excited to the state $|2\rangle$ only if the other qubit is deexcited at the same time. It thus implements an unintentional entangling gate preparing a state of the kind $a |1\rangle_{q2}|0\rangle_{q3}+b|0\rangle_{q2}|2\rangle_{q3}$. To show this, we prepare qubit 2 again in state $|1\rangle$, apply a CNOT gate to qubits 3 and 5, and reset qubit 2 with a Pauli $X$ gate. The results of this circuit are presented in Fig.~\ref{fig_1}(e), thus showing that the populations of qubit 2 and 3 significantly deviate from the expected behaviour and agree with the above entangled state, with $|a|^2\approx 0.65$ and $|b|^2\approx 0.31$. %

In conclusion, the identified frequency collisions lead to severe errors of the involved qubits' outcomes which cannot be explained by the reported average gate and readout errors of the device. 
We note that the strength of the induced errors is in general state dependent, e.g., the above discussed transitions (i.e., Fig.~\ref{fig_1}(a) and (b)) only occur if the first qubit is prepared in the state $|1\rangle$, making it desirable to quantify such errors on average using an appropriate randomized benchmarking protocol.

\section{Crosstalk characterization via randomized benchmarking}\label{sec:CrosstalkChar}

The standard calibration data of \textit{ibmq\_ehningen} involves single- and two-qubit average error rates which are determined using the randomized benchmarking protocol~\cite{IBMQuantum}. The latter is implemented either consecutively on isolated qubits, respectively qubit pairs,  or on well-separated batches of qubits in order to assure that crosstalk between individual characterization experiments is negligible. Hence, the reported average error rates in general underestimate the observed error strength as compared to a situation where gate operations are run on all qubits simultaneously. As a result, any numerical simulation of the system that is based on a noise model involving the reported calibration data will overestimate the resulting circuit fidelity, since crosstalk between qubits is not taken into account. 

\begin{figure}[t!]
\begin{center}
\includegraphics[width=0.48\textwidth]{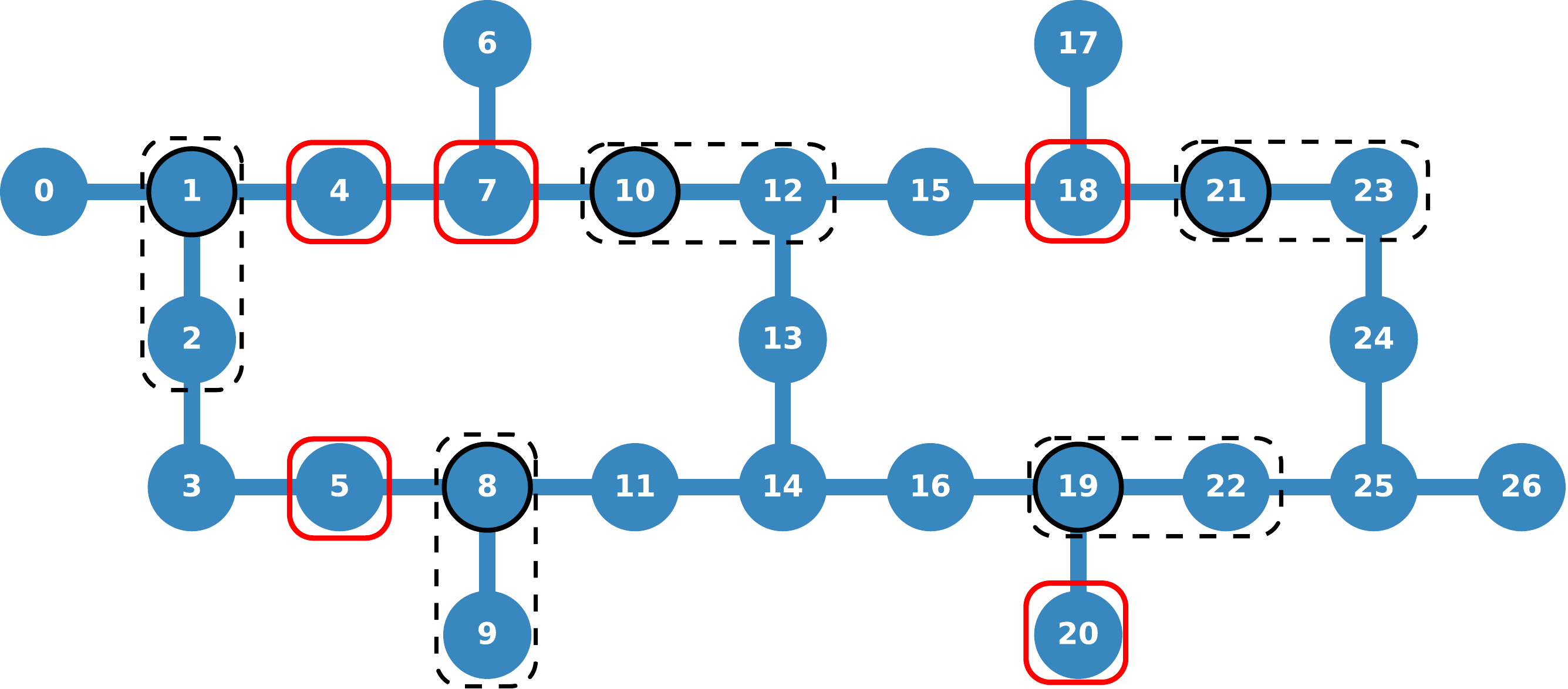}%
\end{center}
\caption{Coupling map of the $27$ qubit Falcon r5.11 transmon processor \textit{ibmq\_ehningen}. Black and red boxes indicate qubit triplets consisting of one qubit pair plus one neighboring qubit, respectively. Black circles indicate those qubits to which a cross-resonance pulse is applied during the implementation of a CNOT gate on the corresponding qubit pair (dashed black box). Shown is one possible characterization batch of in total 12 batches of such triplets (see Appendix~\ref{app:Batches})}
\label{fig_2}
\end{figure}

In order to characterize such crosstalk effects, we use randomized benchmarking run simultaneously on selected sets of qubits, respectively qubit pairs~\cite{Emerson_2005,PhysRevA.77.012307}. To do so, we first analyse the device's properties in order to identify combinations of qubits and qubit pairs which potentially lead to unwanted resonances of the kind discussed in Sec.~\ref{sec:CrosstalkEffects}. That is, for each qubit pair of the  \textit{ibmq\_ehningen} device, we select always one neighbor qubit in the following way. We first inspect the pulse schedule of the respective qubit pair's CNOT gate in order to identify to which of the two qubits the underlying cross-resonance pulse is applied to (see Fig.~\ref{fig_1}(c)) and then select the corresponding adjacent qubits~\cite{Alexander_2020}. In this way, we obtain 41 qubit triplets which we distribute 
onto batches containing non-overlapping sets of triplets that can be characterized in parallel (see Fig.~\ref{fig_2} for an example).

This leads to overall 12 batches that we use to run simultaneous randomized benchmarking (RB) experiments of always one qubit pair in combination with the corresponding neighbor (see App.~\ref{app:Batches} for an example). The RB experiments are designed in such a way that individual Clifford operations are executed approximately simultaneously by inserting circuit barriers over the whole qubit triplet after each Clifford operation. 

\begin{figure}[t!]
\begin{center}
\includegraphics[width=0.48\textwidth]{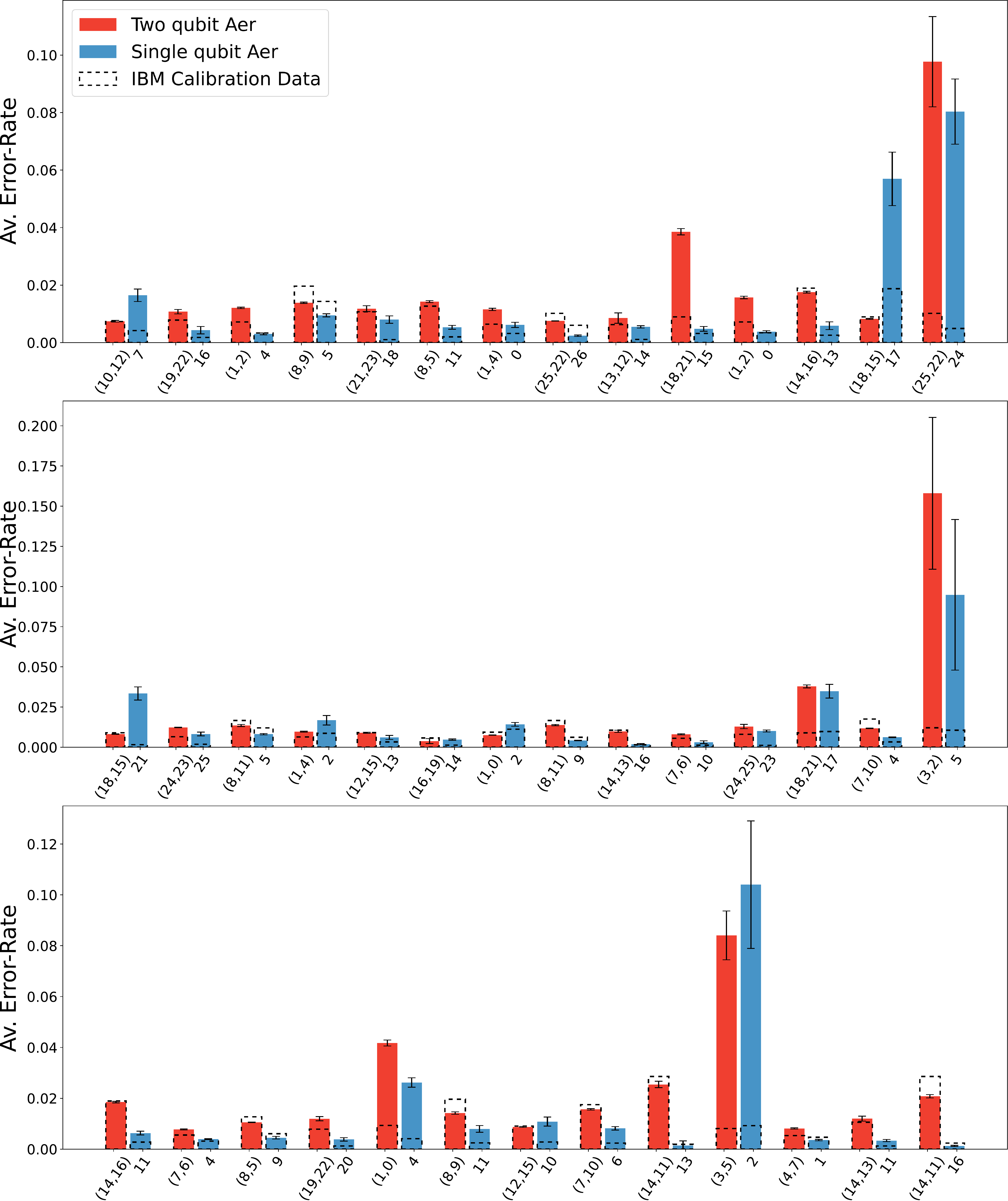}%
\end{center}
\caption{Simultaneously characterized single- (red) and two-qubit (blue) average error rates of 41 three-qubit sets obtained via the randomized benchmarking protocol on \textit{ibmq\_ehningen} (23rd January 2022). The latter was implemented with circuit lengths $[1, 3, 10, 20, 40, 65, 95, 130, 175]$,  each of which was repeated $5$ times with $10\,000$ shots per circuit. Barriers have been inserted into the individual circuits in order to ensure their nearly simultaneous execution. 
Dashed boxes indicate the respective single- and two-qubit average error rates provided by IBM on the same day. 
}
\label{fig_3}
\end{figure}

The results of this characterization experiments are presented in Fig.~\ref{fig_3}. For each qubit triplet, we obtain a pair of one two-qubit and one single-qubit average error rate, which characterize the strength of crosstalk between each other. We find that, in some cases, the simultaneous characterization leads to a significantly larger average error rate as compared to the values reported by IBM~\cite{IBMQuantum}. The latter include the two examples discussed in Sec.~\ref{sec:CrosstalkEffects} but also other previously unknown combinations of qubits. In particular, in some cases, the single-qubit error is increased even though the simultaneously characterized two-qubit error is not. However, one also observes a systematic overestimation of the single-qubit error rates of all triplets as compared to the device's calibration data. This can be explained by the fact that the single-qubit characterization circuits are stretched due to the considerably longer execution time of the simultaneously executed two-qubit circuits. 
As a consequence, the single-qubit circuits involve an unavoidable idle time before the execution of each of the involved Clifford gates which leads to additional decoherence. The latter might be avoided in the future by complementing the single-qubit characterization circuits with dynamical decoupling sequences. 

We emphasize that the above outlined characterization protocol 
leads to reproducible results, which is expected for hardware properties such as frequency collsions. However, we also observe temporary error events which occur mostly on isolated qubits and  lead to
false
readout data, e.g., distorted readout probabilities. In general it is difficult to find the origin of such error events but, in some cases, they 
may be
caused by two-level systems (TLS) which temporarily appear in the hardware and interact with one of the device's qubits.
In other cases, the errors might be due to cosmic radiation which, however, would likely lead to a more drastic breakdown of the device performance~\cite{McEwen:2022ww,2022arXiv221004780T}. 

\section{Simulation of Hardware}\label{sec:CrosstalkNoiseModel}
\subsection{Crosstalk-aware noise model}
In the following, we use the obtained characterization data in order to define an appropriate crosstalk-aware noise model that allows for a more accurate simulation of the respective device.

\begin{figure}[t!]
\begin{center}
\includegraphics[width=0.45\textwidth]{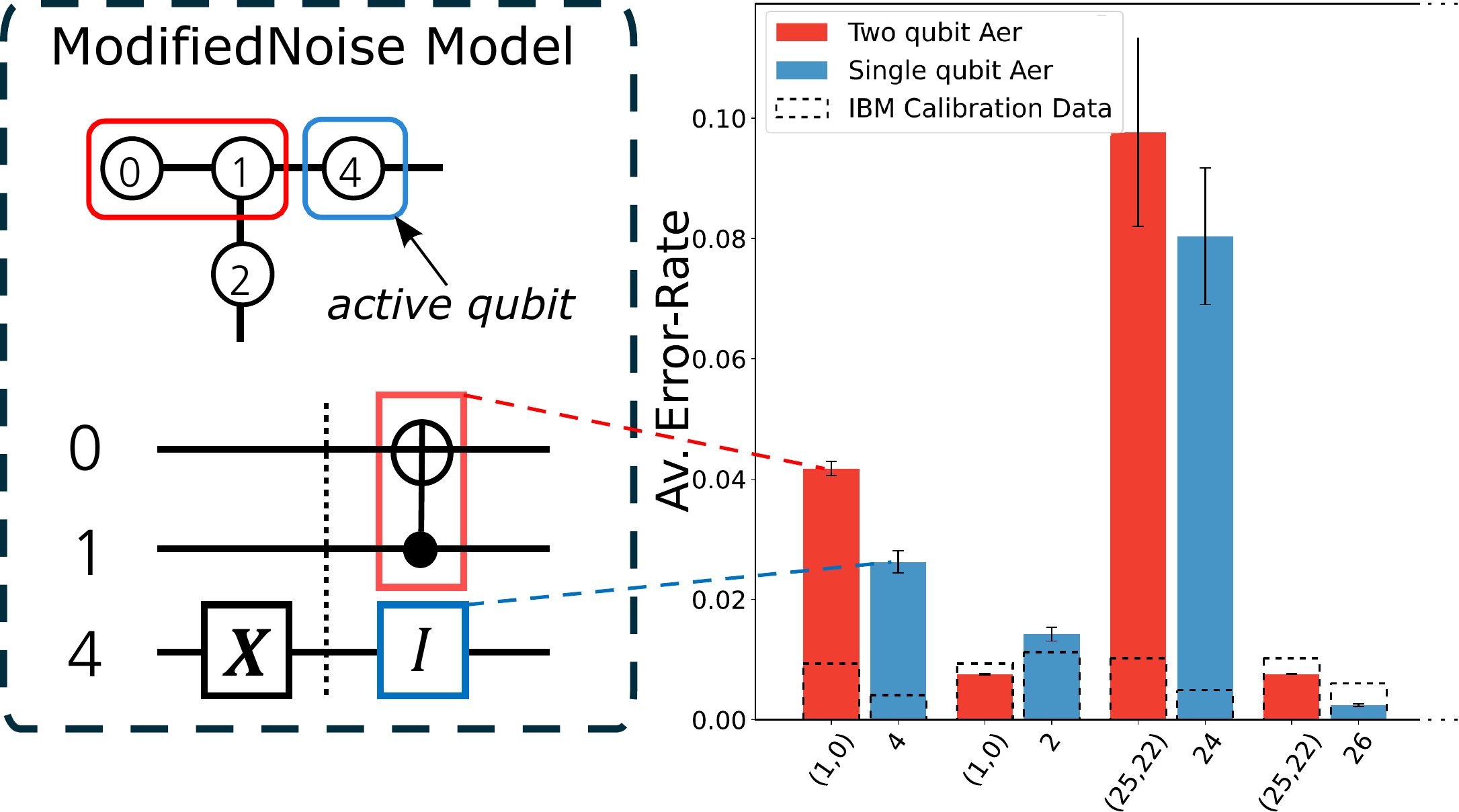}%
\end{center}
\caption{Sketch of the crosstalk-aware noise model. Whenever a CNOT gate is applied to a qubit pair (e.g. $[0,1]$), such that the control qubit is adjacent to another active qubit, we supply it with a depolarization error the strength of which is determined by the characterization data obtained in Sec.~\ref{sec:CrosstalkChar}. At the same time, we attach an additional identity gate with corresponding depolarization error to the adjacent qubit in order to account for the correlations caused by the crosstalk.  }
\label{fig_4}
\end{figure}

Standard noise models as provided, for instance, by the IBM Quantum backends~\cite{IBMQuantum} or as developed in Ref.~\cite{Finsterhoelzl_2023} do not take into account crosstalk between neighboring qubits. They are usually based on the respective devices' calibration data and, as such, complement basis gate operations with amplitude- and phase-damping channels according to the measured single-qubit relaxation and dephasing times, $T_1$ and $T_2$, respectively. Furthermore, depolarizing channels are added to each gate with strengths chosen such that the average error rates of the overall error channels match the measured average error rates reported in the calibration data. Lastly, single-qubit readout errors are added according to the measured probabilities that the recorded classical bits representing the outcomes of measurements performed on the respective qubits are flipped. The simulation results obtained with this type of noise model often deviate from the actual measurement data collected from the hardware (see discussion in Sec.~\ref{sec:CrosstalkNoiseModel} and  results presented in Fig.~\ref{fig_6}).

We modify the standard noise model in order to account for crosstalk effects by following a simple strategy (see Fig.~\ref{fig_4}). First, we scan through the two-qubit gates of the transpiled circuit, i.e., the considered circuit expressed in terms of the basis gate operations of the device, and for each of them check whether the neighboring qubits, according to the triplets defined in Sec.~\ref{sec:CrosstalkChar}, are active. We consider only those qubits as active which, previously to the considered two-qubit gate, have already undergone an arbitrary gate operation. To verify this we analyse the schedule of the respective microwave pulses realizing the circuit's gate operations~\cite{Alexander_2020}. If a neighboring qubit is active we supply the respective two-qubit gate with a depolarizing error with strength given by the two-qubit error rates reported in Fig.~\ref{fig_3}. In case we find that more than one of the neighboring qubits is active, we use the corresponding largest average error rate that has been obtained. If none of the neighboring qubits is active we simply use the average error rates provided by the calibration data of the device~\cite{IBMQuantum}. Furthermore, in order to account for the crosstalk's impact on the active neighbouring qubits, we apply to each of them additional identity operations which are supplied with depolarization errors with strengths determined by the single-qubit average error rates reported in Fig.~\ref{fig_3}.  
Finally,
we extract the idle times of each qubit from the respective pulse schedule of the circuit under consideration and supply the circuit with additional amplitude- and phase-damping channels defined through the  $T_1$ and $T_2$ times reported by the device. 

\subsection{Comparison between simulation and hardware}
\begin{figure}[t!]
\begin{center}
\includegraphics[width=0.49\textwidth]{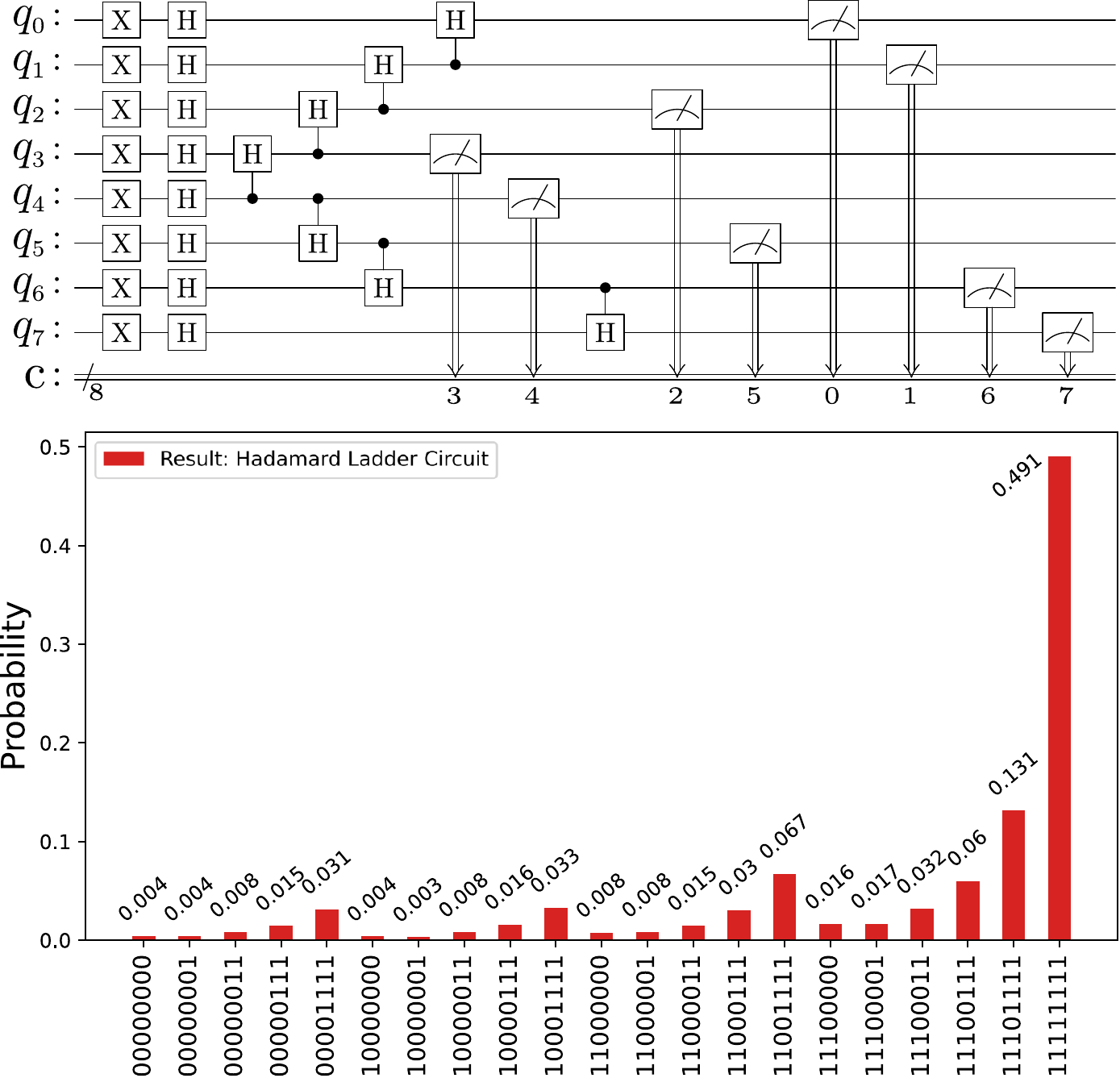}%
\end{center}
\caption{Representation of the 8-qubit Hadamard ladder circuit (top) as well as the exact probability distribution obtained from a computational basis measurement of the error-free output state. }
\label{fig_5}
\end{figure}

To test the above introduced noise models, we use the exemplary Hadamard ladder quantum circuit, depicted in Fig.~\ref{fig_5}, which consists of 8 qubits
and a moderate
number of single- and two-qubit gates.   
The exact probability distribution produced by a computational basis measurement of the Hadamard ladder circuit is presented in Fig.~\ref{fig_5}.

\begin{figure*}[t!]
\begin{center}
\includegraphics[width=0.65\textwidth]{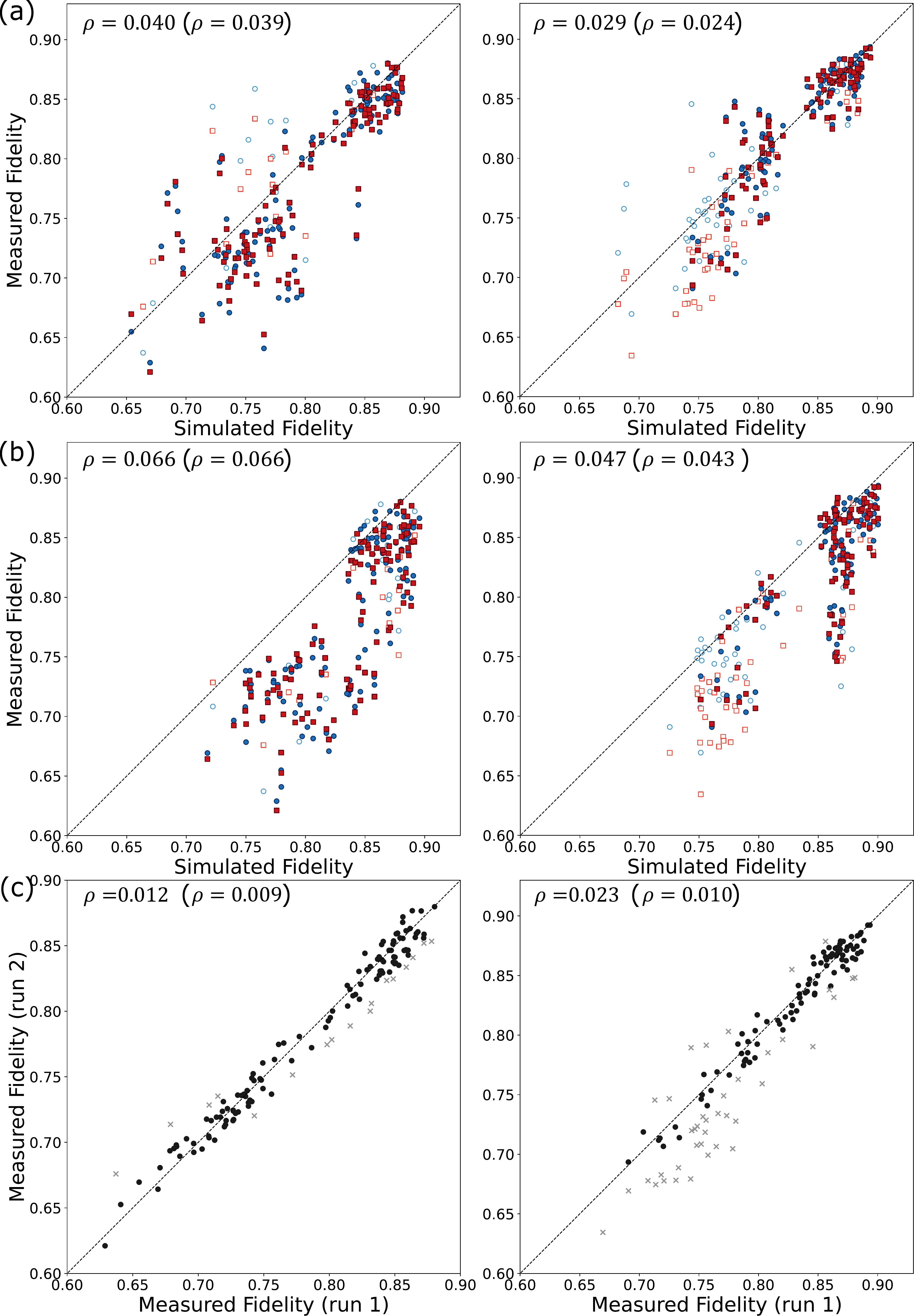}%
\end{center}
\caption{Results of always two consecutive runs of the 8-qubit Hadamard ladder circuit on the Falcon r5.11 processor
\textit{ibmq\_ehningen}
on the 23rd (left column) and 24th (right column) January 2023 for all 132 possible initial qubit layouts. (a) Correlations between the measured Hellinger fidelities and those obtained by our crosstalk-aware noise model, where  the characterization has been performed in between run 1 (circles) and 2 (squares). (b) Correlations between the measured Hellinger fidelities and those obtained by the standard noise model constructed from the backend calibration data. (c) Correlations between the measured Hellinger fidelities of run 1 and 2 for all 132 possible initial qubit layouts. Light crosses indicate those layouts where the fidelity between run 1 and 2 deviate more than $2\%$. The same layouts are indicated in (a,b) by light blue empty circles, respectively, red squares. $\rho$ indicates the root-mean-square deviation between each of the compared data sets, while the values in parenthesis refer to the reduced data sets. In all cases, the crosstalk-aware noise model (a) exhibits a smaller deviation from the data measured on \textit{ibmq\_ehningen} than the standard noise model (b).}
\label{fig_6}
\end{figure*}

Before running the above circuit either on the simulator or on the \textit{ibmq\_ehningen} device, we first determine all possible qubit layouts of the $27$-qubit topology of the Falcon r5.11 processor. Doing so is easy for the Hadamard ladder circuit as it involves only two-qubit gates applied to nearest-neighbor qubits, which can be mapped to chain of qubits. Hence, it remains to simply count all possible 
$8$-qubit chains that can be supported by the topology, leading to overall $132$ different qubit layouts. The latter can now be used to run the Hadamard ladder circuit on \textit{ibmq\_ehningen} and compare the outcomes with results obtained by the simulator using the introduced noise models. In order to investigate the reproducibility of the  results we perform each run twice in a row and implement the device characterization (see Sec.~\ref{sec:CrosstalkChar}) in between these two runs. 
Lastly, we also repeat the whole procedure on different days to investigate possible long term effects. 

We note that the simulations in terms of the discussed noise models represent average case scenarios, since the supplied noise channels are mainly based on a set of average error rates. Hence, to take into account that the noise of a specific circuit might deviate from this average case we modify the circuits that are ran on the real hardware with appropriate randomized compilings~\cite{PhysRevA.94.052325,PhysRevX.11.041039}. The latter modify the device's noise in such a way that it is 
transformed into
so-called Pauli noise, i.e., noise that can be described by a quantum channel that is diagonal in the Pauli basis. To do so, we insert into the circuits under consideration additional randomly sampled single-qubit Pauli gates which leave the noise-free realization of the circuit unchanged. As the main source of errors, according to our assumptions, occurs within the qubit triplets introduced in Sec.~\ref{sec:CrosstalkChar} we insert these gates always before and after each CNOT gate, as well as its 
active
neighbors. In this way, we produce 100 randomly compiled circuits for each layout, run them with 100 shots each and combine the obtained results in post-processing. In contrast, on the simulator we run each layout once with $10'000$ shots. In case of the hardware runs, we switched off all error suppression and mitigation options that are provided by the backend~\cite{IBMQuantum,PRXQuantum.2.040326,PhysRevLett.119.180509,Kandala:2019tv,https://doi.org/10.48550/arxiv.2201.09866}.

We use the so-called Hellinger fidelity 
\begin{align}
    F(Q,P)=\left(\sum_i \sqrt{p_i q_i}\right)^2,
\end{align}
where $Q=\{q_i\}$ and $P=\{p_i\}$ denote discrete probability distributions, 
to gauge how close the observed measurement distributions are to the error-free distribution presented in Fig.~\ref{fig_5}. 
Figures~\ref{fig_6}(a-c) thus present the comparisons between the measured Hellinger fidelities and those obtained from simulations with the two noise models under consideration, i.e., our crosstalk-aware noise model and the standard one obtained from the backend's calibration data. These results show that taking into account crosstalk effects into the noise model leads to a quantitatively better description of the hardware results, i.e., to an improvement of the root-mean-square deviation $\rho$ between the simulated and the measured data, see Fig.~\ref{fig_6}(a). In contrast, the backend noise model generally underestimates the errors of the device and thus leads to a systematic overestimation of the fidelities, see Fig.~\ref{fig_6}(b).  
The improvement of the crosstalk-aware noise model  becomes most distinct for those layouts which yield  measured fidelities below roughly $0.85$, while for layouts with fidelities above this threshold crosstalk effects do not play a major role and thus both noise models perform similarly well.

While improved, the predictions of the noise models are still quite dispersed, which can be partially attributed to short term fluctuations of the hardware results. We analyse the latter in Fig.~\ref{fig_6}(c), where we compare the Hellinger fidelities of the two consecutive hardware runs for each day, respectively. We observe a certain number of outliers, for instance, in case of the runs performed on 
the 24th January 2023 (right column).
They originate from
fluctuations occurring over the 
time scale during which the circuits are run.  
Such fluctuations (possibly due to temporary TLS defects, as mentioned in the introduction) 
lead to irreproducible results and are beyond the reach of our crosstalk-aware noise model.
Hence, if we focus on those layouts where these fluctuations are small, e.g., the difference of fidelities between both consecutive runs smaller than $0.02$, the description of the measured data in terms of our noise model should improve. The latter is confirmed if we compare the measured Hellinger fidelities and the simulated ones only for those layouts, see the black symbols in Figs.\ref{fig_6}(a,b)).

\section{Conclusions and Outlook}\label{sec:Conclusions}

We presented a detailed analysis of crosstalk effects 
occurring on fixed-frequency superconducting transmon processors. We first discussed the physical origin of crosstalk effects caused by unwanted resonances resulting from frequency collisions on the device and showed how they manifest themselves in strong correlations between simultaneously executed gate operations. We then proceeded to characterize the magnitude of these effects using a simultaneous randomized benchmarking procedure, where combinations of single- and two-qubit gates are characterized simultaneously on selected subsets of neighboring qubits. Conducting this characterization on all designated qubit neighbors over the whole device provided us with a detailed map of crosstalk-prone combinations of qubit pairs with adjacent qubits. Finally, we used this information in order to devise a crosstalk-aware noise model and demonstrated its improved performance compared to standard noise model based on the backend's calibration data alone. 

Characterizing and modelling crosstalk of superconducting transmon processors is an important task for the development of NISQ-era quantum algorithms. On the one hand, such an understanding allows one to develop more accurate means to simulate current quantum computing hardware and thus to make more realistic predictions of the performance of various quantum computing algorithms without the need of using the real hardware. On the other hand, characterizing the devices' errors is the first step towards developing novel and better error suppression and mitigation protocols.

Lastly, we emphasize that the present characterization scheme is efficiently implementable also on larger transmon architectures involving more qubits than the presently studied $27$-qubit device. Moreover, slight adjustments to the present scheme enable crosstalk investigations also on other hardware platforms based on, e.g., trapped ions or nitrogen vacancy centers. In this case, however, the method of choosing designated qubit subsets will differ due to the hardware-specific realizations of individual gate operations.

\begin{acknowledgments}
The authors acknowledge funding from the  
Ministry of Economic
Affairs, Labour and Tourism Baden-Württemberg, under the
project QORA in the frame of the Competence Center Quantum Computing Baden-Württemberg.

\end{acknowledgments}

\appendix

\section{Kraus representations of noise channels}

\subsection{Thermal Relaxation Error}
The interaction of the qubits with the environment can be modeled by two contributions. On the one hand, the qubits undergo thermal decay which can be modelled by an amplitude damping channel the Kraus operators of which read as follows (for a single qubit):
\begin{eqnarray}
E^{\rm D}_{1} &=& \begin{pmatrix}
 1 & 0 \\
 0 & \sqrt{1-p_\text{AD}} \\
 \end{pmatrix},\   
 E^{\rm D}_{2} =\begin{pmatrix}
 0 & \sqrt{p_\text{AD}} \\
 0 & 0 
 \end{pmatrix}, 
\end{eqnarray}
where $p_\text{AD}$ describes the probability of 
decay from state $|1\rangle$ to state $|0\rangle$.

On the other hand, dephasing 
is modelled by the following Kraus operators:
\begin{eqnarray}
E^{\rm PD}_{1} &=& \begin{pmatrix}
 1 & 0 \\
 0 & \sqrt{1-p_\text{PD}} \\
 \end{pmatrix},\   
 E^{\rm PD}_{2} =\begin{pmatrix}
 0 & 0 \\
 0 & \sqrt{p_\text{PD}} 
 \end{pmatrix}. 
\end{eqnarray}
Combining the two processes leads to the following Kraus representation~\cite{PhysRevA.86.062318}:
\begin{eqnarray}
E^{\rm D}_{1} &=& \begin{pmatrix}
 1 & 0 \\
 0 & \sqrt{1-\gamma - \lambda} \\
 \end{pmatrix} \nonumber \\ &=& \frac{1+\sqrt{1-\gamma - \lambda}}{2}\mathbb{I} + \frac{1-\sqrt{1-\gamma - \lambda}}{2}\sigma^{\rm z}, \nonumber \\ 
 E^{\rm D}_{2} &=& \begin{pmatrix}
 0 & \sqrt{\gamma} \\
 0 & 0 \\
 \end{pmatrix} = \frac{\sqrt{\gamma}}{2}\sigma^{\rm x}+\frac{i\sqrt{\gamma}}{2}\sigma^{\rm y}, \nonumber \\
E^{\rm D}_{3} &=& \begin{pmatrix}
 0 & 0 \\
 0 & \sqrt{\lambda} \\
 \end{pmatrix}=\frac{\sqrt{\lambda}}{2}\mathbb{I}-\frac{\sqrt{\lambda}}{2}\sigma^{\rm z},
\label{eq:xdef}
\end{eqnarray}
where, $\gamma \equiv p_{\rm AD}$ and $\lambda \equiv (1-p_{\rm AD})p_{\rm PD}$, and
\begin{eqnarray}
&1-p_{\rm AD} = e^{-t/T_{\rm 1}}, & \\
&\sqrt{\left(1-p_{\rm AD}\right)\left(1-p_{\rm PD}\right)} =e^{-t/T_{\rm 2}}. &
\end{eqnarray}
with the thermal relaxation and decoherence times $T_1$ and $T_2$.

\subsection{Depolarization Error}
A simplified error model is provided by the depolarization channel which assumes that the qubit undergoes with probability $p$ a certain Pauli error, i.e., either a bit-flip, phase-flip, or both. The corresponding channel's Kraus representation reads:
\begin{align}
    \mathcal E (\rho)=(1-3p)\rho +p (X\rho X+Y\rho Y+Z\rho Z).
\end{align}

\section{Example of batches of qubit triplets}\label{app:Batches}
In the following, we show an example of the 41 qubit triplets distributed over 12 batches such that the individual triplets are well separated on the qubit topology shown in Fig.~\ref{fig_2}:
\begin{enumerate}
    \item \ ([18, 21], [15]), ([14, 13], [11]), ([24, 25], [23]), ([1, 4], [2]), ([10, 12], [7])
    \item \ ([25, 22], [26]), ([1, 2], [0]), ([18, 17], [21]), ([14, 16], [13])
    \item \ ([8, 5], [11]), ([19, 22], [20]), ([18, 15], [21]), ([1, 4], [0]), ([24, 23], [25])
    \item \ ([1, 2], [4]), ([14, 16], [11]), ([18, 21], [17]), ([25, 22], [24])
    \item \ ([4, 7], [1]), ([21, 23], [18]), ([14, 13], [16]), ([3, 2], [5])
    \item \ ([8, 9], [11]), ([7, 6], [4]), ([18, 17], [15]), ([13, 12], [14])
    \item \ ([1, 0], [4]), ([12, 15], [10]), ([16, 19], [14]), ([8, 9], [5])
    \item \ ([8, 11], [5]), ([7, 10], [4]), ([18, 15], [17]), ([19, 22], [16])
    \item \ ([1, 0], [2]), ([12, 15], [13]), ([8, 5], [9])
    \item \ ([7, 6], [10]), ([3, 5], [2]), ([14, 11], [16])
    \item \ ([14, 11], [13]), ([7, 10], [6])
    \item \ ([8, 11],[9])
\end{enumerate}

\bibliographystyle{apsrev4-2}
\bibliography{refs}

\end{document}